\documentclass[showpacs,aps,graphicx,twocolumn]{revtex4}

\usepackage{graphicx}
\usepackage{amssymb}
\usepackage{amsmath}
\usepackage{subfigure}

\begin{document}

\title{The second order magnetic field gradient induced strong coupling between nitrogen-vacancy centers and a mechanical oscillator}
\author{Kang Cai$^1$, Rui-Xia Wang$^1$, Zhang-Qi Yin$^2$\footnote{{Email address:
yinzhangqi@tsinghua.edu.cn}}, Gui-Lu Long$^{1,3,4,}$\footnote{{Email address:
gllong@tsinghua.edu.cn}}}
\affiliation{
$^1$ State Key Laboratory of Low-Dimensional Quantum Physics and Department of Physics, Tsinghua University, Beijing 100084, China\\
$^2$ Center for Quantum Information, Institute for Interdisciplinary Information Sciences, Tsinghua University, Beijing 100084, China\\
$^3$ Tsinghua National Laboratory of Information Science and Technology, Beijing 100084, China\\
$^4$ Collaborative Innovation Center of Quantum Matter, Beijing 100084, China}

\begin{abstract}
We consider a cantilever mechanical oscillator(MO) made of diamond. There is a nitrogen-vacancy(NV) center at the
end of the cantilever. Two magnetic tips induce strong second order magnetic field gradient near the NV center.
Under a coherent driving on MO, we find that the coupling between the MO and the NV center can be greatly enhanced.
We studied how to realize quantum state transfer between MO and NV center and generate entanglement between them.
We also proposed a scheme to generate the two-mode squeezing between different MO modes by coupling them to the same NV center.
The decoherence and dissipation effects for both MO and NV center are numerically calculated by taking the present
experimental parameters. It is found that high fidelity quantum state transfer, entanglement generation, and large two-mode squeezing
could be achieved.
\end{abstract}
 \maketitle

\section{introduction}

Nano(Micro)-mechanical oscillator, due to its applications in ultra-high precise sensing and testing quantum phenomenon at macroscopic scale \cite{Marquardt,Yinreview}, has
attracted a lot of attentions in recent years. Combined with cavity opto- and
electro-mechanics, the mechanical oscillator (MO) has been explored extensively as a quantum interface \cite{Duan,Tian,Andrews2014,Wang2012}. Besides, strong
coupling for hybrid MO systems have been realized, which can be used to achieve ground state cooling, quantum information processing etc \cite{Kippenberg2007,HLWang2009,YFXiao2015,Painter2015}.
 Recently, more and more attentions have been paid on
interfacing the mechanical degrees of freedom with a single quantum object
such as a 2-level system whose quantum state can be precisely controlled.
It has been investigated both theoretically and experimentally of coupling a nanomechanical oscillator
with solid state qubits, such as Nitrogen-vacancy (NV) centers in diamond \cite{Rugar,Rabl2009,Xu,XBwang,Rabl2010,PengboLi2015,PengboLi2016,Connell,Arcizet,Lukinscience2012,Neukirch2015,Hoang2016}.

Resonating nanostructures made
of single-crystal diamond are expected to possess excellent mechanical
properties, including high-quality factors and low dissipation. Diamond has been expected to have great applications as a uniquely versatile material, yet one
that is intricate to grow and process. Fortunately,
quality factors exceeding one million are found at room temperature,
surpassing those of single-crystal silicon cantilevers of similar dimensions
by roughly an order of magnitude \cite{Tao}. In addition, diamond hosts interesting
intrinsic dopants \cite{HainlinWang}---most prominently the NV center---that have been
recognized as rich resource for single-photon generation, quantum
engineering and nanoscale magnetic sensing. With all these advantages, the
diamond nanostructures may lead to wide applications in quantum science.

NV centers are formed by a nitrogen atom and a nearby vacancy in diamond, usually negatively charged, possessing 6 electrons, with spin
$S=1$ in the ground state and regarded as artificial atoms in solid systems \cite{Doherty2013}.
 Because of the long coherence time and sensitive to magnetic field,
they are promising candidates for quantum information processing and
also widely used as solid-state ultra sensitive magnetic field sensor. There
are usually two kinds of methods to couple the NV centers with MO. The first one
requires the strain induced effective electric field to mix phonon mode with
NV centers electron spins \cite{Lukin2013,Teissier2014,Yin2015}. The strain induced
coupling is very sensitive to the size of the MO. The strong coupling regime is
very difficult to approach through this direction.
The second one is based on the strong first order magnetic field gradient \cite{Rabl2009}.
For the first method, one of the main problems in the hybrid systems of NV center and
MO is the strong coupling condition requiring ultra-high
magnetic gradient \cite{Rabl2009}. One way to solve the problem is
to reduce the effective mass of the oscillator or trapping frequency \cite{Yin2013,2016WCGeNJP}.
Here we study the third coupling mechanism between NV
center and MO based on second order magnetic
coupling, which was studied for heating one mode of the MO to cool the other one to quantum
ground regime \cite{Yin2016}.

In this article, we propose a scheme to realize strong coupling between MO and NV center under the
second order magnetic field gradient.
In section II, we firstly introduce a new model used to describe the
coupling between MO and NV centers and demonstrate the increased coupling
between them. In section III, we propose the scheme to generate the
entanglement and realize state transfer between the NV center and the MO. Then we generalize our model to the interaction between
the MO and the NV centers ensemble. In section IV, we show another
application in two-mode squeezing of the MO. In section V, we give the brief discussion and
summary.

\section{Model}

\begin{figure}[htbt]
\begin{center}
\includegraphics[width=6cm,angle=0]{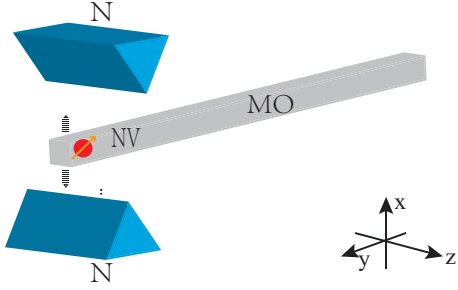}
\caption{(Color online) The model of the second order magnetic field gradient coupling between NV center and diamond MO. NV centers are located at the end of MO. Two magnetic tips are symmetrically placed on both sides of MO to construct the second magnetic field gradient near NV centers. With the external driving, the MO oscillates along x direction. A homogeneous external magnetic field is along z direction. }\label{Model}
\end{center}
\end{figure}

As shown in Fig.\ref{Model}, a nano-diamond MO is fabricated and
oscillates along the $x$ axis with two modes $\omega _{a}$ and $\omega _{b}$%
, while one NV center is hosted at the end of the oscillator \cite{Yin2016}. Two
magnetic tips point at each other along $x$ axis and are situated at two
sides of oscillator symmetrically. External static magnetic field $B_{ext}$
is added along $z$ axis. The frequency difference between NV center electron spin states
$|-1\rangle$ and $0\rangle$ is defined as $\omega _{z}$. We denote the $|0\rangle$ as ground state $|g\rangle$, and the $-1\rangle$
as the excited $|e\rangle$. The driven
force is exerted on the oscillator with frequency $\omega _{a}$. The
energy split of NV center electron spin is tuned to satisfy the relation $\omega _{z}=\omega _{b}-\omega _{a}=\Delta$.
In the rotating frame $H_{10}=\omega
_{a}a^{\dagger }a+\omega _{a}b^{\dagger }b$, the two modes Hamiltonian can
be simplified as \cite{Yin2009,Yin2016}%
\begin{equation}
\begin{aligned}
H_{0} &=\frac{\omega _{z}}{2}\sigma _{z}+\triangle b^{\dagger }b+\frac{%
\Omega _{1}}{2}\left( a+a^{\dagger }\right) +\frac{\Omega _{2}}{2}\left(
b+b^{\dagger }\right),  \\
H_{1} &=\left[ g_{a}a^{\dagger }a+g_{b}b^{\dagger }b+g_{ab}\left(
a^{\dagger }b+b^{\dagger }a\right) \right] \sigma _{x},
\end{aligned}
\end{equation}%
where $a\left( a^{\dagger }\right) $, $b\left( b^{\dagger }\right) $
represent the annihilation (creation) operator of two oscillator phonon
modes, respectively, $\Omega _{1}\left( \Omega _{2}\right) $ is the driven
strength for the mode $a\left( b\right) $, $g_{a}=g_{b}=g_{ab}=g$ are
coupling strength with second order magnetic field gradient. The Pauli operators
are defined as $\sigma_z = |e\rangle\langle e| - |g\rangle \langle g|$, $\sigma_x =
|e\rangle \langle g| + |g\rangle \langle e|$. Due to the
symmetrical location of two magnetic tips, the first order coupling
interaction between NV center and MO vanishes and the second order
interaction begins to play the essential part which is described by the $%
H_{1}$.

Firstly, we initialize the NV center to the ground state, and then only the
uncoupling Hamiltonian term $H_{0}$ matters. In the presence of the driving,
 the MO tends to behave coherently. The dynamics of the
MO can be derived through quantum Langevin equation \cite{Scully}%
\begin{equation}
\begin{aligned}
\dot{a} &=-i\left[ a,H_{0}\right] -\frac{\gamma _{1}}{2}a+\sqrt{\gamma _{1}}%
a_{in}, \\
\dot{b} &=-i\left[ b,H_{0}\right] -\frac{\gamma _{2}}{2}b+\sqrt{\gamma _{2}}%
b_{in},
\end{aligned}
\end{equation}%
where $\gamma _{1},$ $\gamma _{2}$ are dissipation rates of modes $a$ and $b$, and $%
a_{in},$ $b_{in}$ are output noise operator of modes $a$ and $b$, with $\langle a_{in}\rangle
=\langle b_{in}\rangle =0$. The steady
state amplitude of each mode satisfies the relations as below%
\begin{equation}\label{amplitude}
\begin{aligned}
i\frac{\Omega _{1}}{2}+\frac{\gamma _{1}}{2}\alpha  &=0, \\
i\triangle \beta +i\frac{\Omega _{2}}{2}+\frac{\gamma _{2}}{2}\beta  &=0,
\end{aligned}
\end{equation}
where $\alpha =-i\Omega _{1}/\gamma _{1}$ and $\beta =-\Omega _{2}/2\Delta
$.  We can see that by increasing the amplitude of
phonon mode and thus, increases the coupling strength between NV center and MO.
Once the steady state established, we turn the NV center
into excited state and the second order interaction between the NV center
and MO comes into effect. Near steady state, we can make transformation
as $a\rightarrow a+\alpha ,b\rightarrow b+\beta$.
The Hamiltonian near steady state takes the form%
\begin{equation}
\begin{aligned}
H_{0} =&\frac{\omega _{z}}{2}\sigma _{z}+\triangle b^{\dagger }b, \\
H_{1} =&\left[ g\left( a^{\dagger }+\alpha ^{\ast }\right) \left( a+\alpha
\right) +g\left( b^{\dagger }+\beta ^{\ast }\right) \left( b+\beta \right)
\right. \\
&\left. +g\left( a^{\dagger }+\alpha ^{\ast }\right) \left( b+\beta \right)
+g\left( b^{\dagger }+\beta ^{\ast }\right) \left( a+\alpha \right) \right]
\sigma _{x},
\end{aligned}
\end{equation}%
In rotating frame $H_{0}=\frac{\omega _{z}}{2}\sigma _{z}+\triangle
b^{\dagger }b$ and rotating frequency approximation, the
effective Hamiltonian can be got as
\begin{equation} \label{eq:HE}
H_{E}=g\left( \alpha ^{\ast }b\sigma ^{+}+\alpha b^{\dagger }\sigma
^{\_}\right).
\end{equation}
where $\sigma^+ = |e\rangle \langle g|$ and $\sigma^- = |g\rangle \langle e|$.

We set the frequency of NV center as $\omega _{z}/2\pi =10MHz$, the mode
frequency of MO $\omega _{a}/2\pi =20MHz,$ $\omega _{b}/2\pi =30MHz$,
driving frequency $\omega _{L}/2\pi =\omega _{a}/2\pi =20MHz$, $\Omega _{1}/2\pi =1.25MHz,$ the dissipation of the
a mode $\gamma _{1}/2\pi =25Hz$. The second order magnetic field gradient could be in the
order of $10^{14} - 10^{15}~ T/m$, corresponding to coupling $g/2\pi \sim 1-10$ Hz \cite{Yin2016}. Here we choose the coupling
strength $g/2\pi =5Hz$. We get $\left\vert \alpha \right\vert =%
\frac{\Omega _{1}}{\gamma _{1}}=50000$. Thus, the effective coupling strength
can be enhanced by $50000$ times. As $\left\vert \alpha \right\vert \gg 1$,
the effective coupling strength between the NV center and the MO is greatly enhanced.

Furthermore, if there are many NV centers in the end of the MO, the effective interaction between the MO
and NV centers could be increased by $\sqrt{N}$, where $N$ is the number of NV centers \cite{Yin2007}. Following the above steps,
we can derive the effective Hamiltonian as%
\begin{equation}
H_{NE}=g\left( \alpha ^{\ast }dJ^{+}+\alpha d^{\dagger }J^{\_}\right),
\end{equation}%
where $J^{+}=\sum_{i=1}^{N}\sigma _{i}^{+}$ and the interaction among NV
centers has been neglected. For this system, from the following analysis,
the interaction time decreases to $t/\sqrt{N}$ and the fidelity of the state
has been lifted.

\section{Entanglement and State Transfer}

The basic requirement of quantum information processing is entanglement and state transfer between the NV center and MO.
In this section, we will discuss the entanglement and state transfer
for the system of NV center and diamond MO.

We assume that the MO is cooled to near the ground state and NV center is in the state $|e\rangle$.
We neglect the effects of decoherence at first. From Eq. \eqref{eq:HE} the system evolves as%
\begin{equation}\label{eq:wave}
|\psi \left( t\right) \rangle =-\sin \left( g\left\vert \alpha \right\vert
t\right) |1,g\rangle +\cos \left( g\left\vert \alpha \right\vert t\right)
|0,e\rangle ,
\end{equation}%
where $|n,g\rangle $ $\left( |n,e\rangle \right) $ represent the system
state with MO in Fock state $|n\rangle $ and NV\ center
in ground (excited) state, respectively.

From the wave function Eq. \eqref{eq:wave}, we can see that at time $t=\pi /4g\left\vert \alpha
\right\vert $, entanglement between NV center and MO is maximal. The
entanglement of system can be measured through negativity, which defined as $%
N\left( \rho \right) =\frac{\left\vert \left\vert \rho ^{T_{A}}\right\vert
\right\vert -1}{2}$ \cite{Peres,Vidal2002}, where the partial transformed density matrix $%
\rho ^{T_{A}}$ has elements $\rho _{m\mu ,n\nu }^{T_{A}}=\rho _{n\mu ,m\nu }$
and the trace norm $\left\vert \left\vert \rho ^{T_{A}}\right\vert
\right\vert =tr\sqrt{\rho ^{T_{A}\dagger }\rho ^{T_{A}}},$ which corresponds
to the sum of the absolute value of eigenvalues of $\rho ^{T_{A}}.$ At this
situation, the negativity can be obtained as $N\left( \rho \right)
=\left\vert \sin \left( 2g\left\vert \alpha \right\vert t\right) \right\vert
/2$. The maximal value of $N\left( \rho \right) $ is $1/2$, corresponding to
the maximal entanglement.

The JC Hamiltonian Eq. \eqref{eq:HE} can also be used for quantum state
transfer between NV center and the MO.
As the MO's oscillation begins, the interaction between the
MO\ and NV\ center is generated and results in the energy exchange between
MO\ and NV\ center. We can see that at $t=\pi /2g\left\vert \alpha
\right\vert $, the state is a product state and the state transfer has been
realized, from $|0,e\rangle $ to $|1,g\rangle $ and at this moment, the
state entanglement decreases to zero corresponding to the minimal value of $%
N\left( \rho \right) $.

In experiment, the effect of the environment should be considered and the
state of the system is usually in the mixed state, expressed by density
matrix $\rho \left( t\right) $. We use the following master equation to describe the
evolution of the density matrix
\begin{equation}
\begin{aligned}
\dot{\rho}\left( t\right)=&-i\left[ H_{E},\rho \left( t\right) \right]
+\frac{\gamma _{2}}{2}\left( \overline{n}_{T}+1\right) L\left[ b\right]
\rho \left( t\right)\\
&+\frac{\gamma _{2}}{2}\overline{n}_{T}L\left[
b^{\dagger }\right] \rho \left( t\right)+\frac{d }{2}L\left[ \sigma ^{z}\right] \rho \left( t\right),
\end{aligned}
\end{equation}%
where $\gamma _{2},d $ denote the dissipation rate of MO and
dephasing rate of NV center, respectively and $L\left[ o\right] \rho =2o\rho
o^{\dagger }-o^{\dagger }o\rho -\rho o^{\dagger }o$. In realistic
experiment, the initial state of phonon is in thermal state and have the
form as $\rho _{0}=\sum_{n=0}^{\infty }p_{n}|n\rangle \langle n|$ with $p_{n}=\langle n\rangle ^{n}/\left( 1+\langle
n\rangle \right) ^{n+1}$ where $\langle n\rangle =\langle b^{\dagger
}b\rangle $ is the mean thermal phonon number of MO. Here, we suppose
MO has been cooled near ground state. The initial state
of the system is $\rho \left( 0\right) =\left\vert e\rangle \langle
e\right\vert \otimes \rho _{0}$ and the space we choose is $\left\{
|g\rangle ,|e\rangle \right\} \otimes \left\{ |n\rangle \right\}
_{n=0}^{n_{\max }}$, with the upper cutoff $n_{\max }=20.$ Set the
dissipation rate of MO $\gamma _{2}=5g$ and dephasing rate of NV center $%
d =50g$ and neglect the spontaneous decay rate of NV center. Put
this system in the low temperature environment with $\overline{n}_{T}=20$
which corresponds to the temperature around $T\simeq 10mk.$

\begin{figure}[htbp]
\begin{minipage}[t]{\linewidth}
  \includegraphics[width=7cm,angle=0]{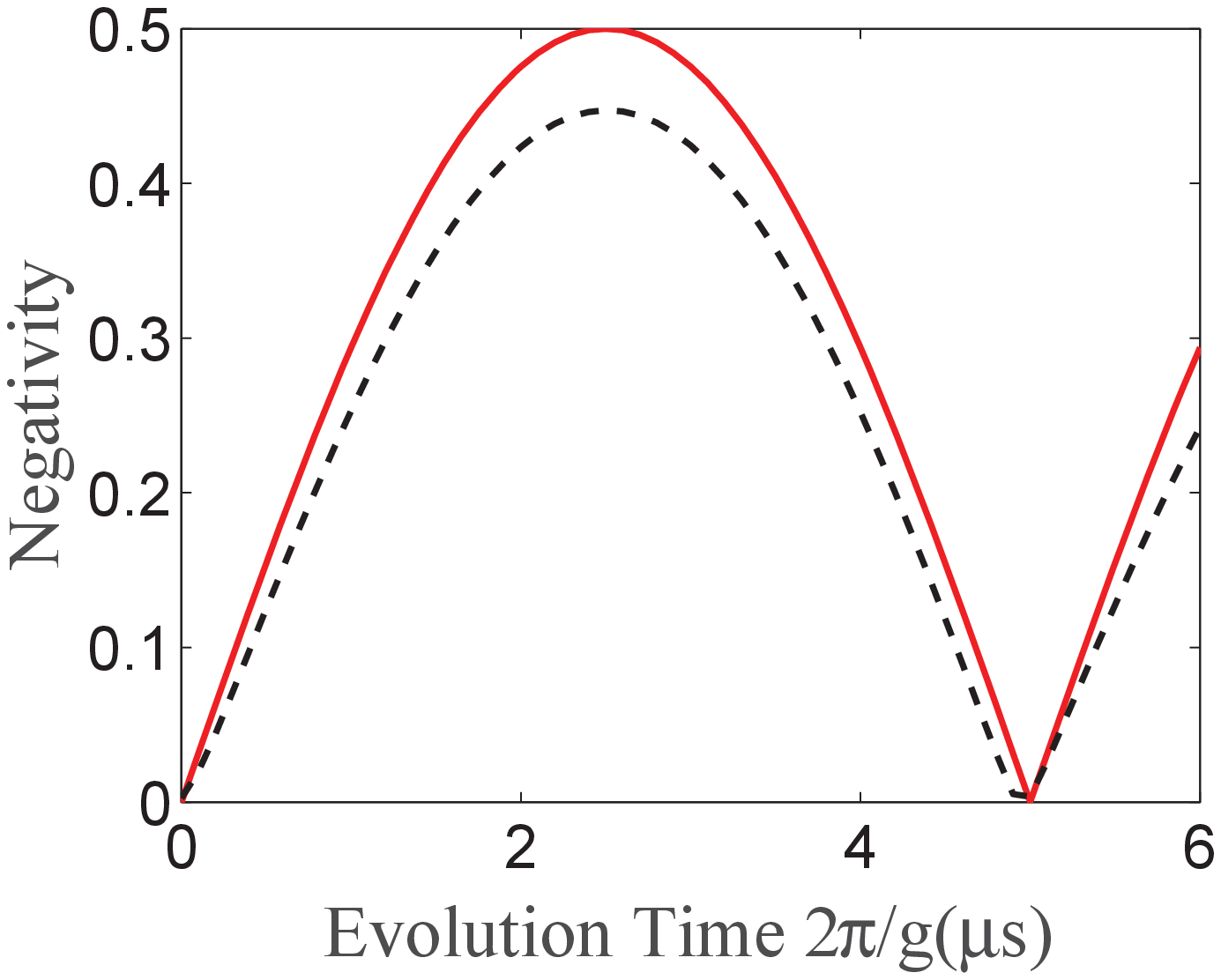}
\end{minipage}
\begin{minipage}[t]{\linewidth}
  \includegraphics[width=7cm,angle=0]{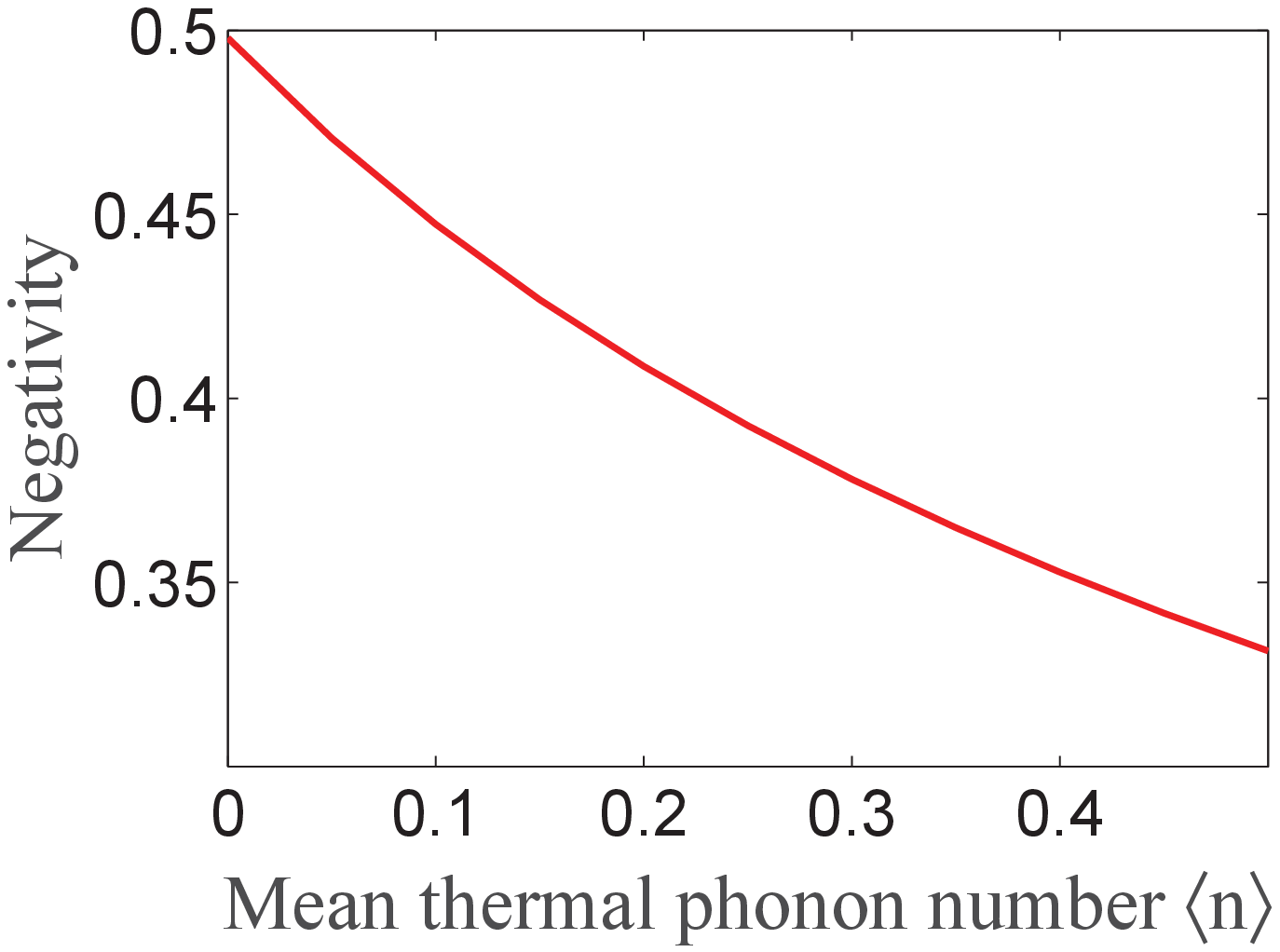}

\end{minipage}
\caption{(Color online) Entanglement between NV center and MO. In upper figure, the red line is the evolution of entanglement for pure state while the black dash line denotes the entanglement in realistic experiment with thermal phonon number $\langle n\rangle=0.1$. The lower figure shows the relation between negativity of maximal entanglement and the mean phonon number of MO. The coupling strength is $g/2\pi=5Hz$, the driving strength is $\Omega/g=2.5\times 10^6$ and the dissipation rate of MO and dephasing rate of NV center are $\gamma/g =5$ and $d/g =50$, respectively. }
 \label{entanglement}
\end{figure}

Once the interaction begins, energy exchange happens between NV center and
MO and consequently, entanglement rises up. Here, we explore the
entanglement of system using negativity, which can be read from Fig.\ref{entanglement}.
The red line depicts the negativity of pure state and the values of
negativity increase with the increasing of entanglement. The maximal and
the minimal values are $1/2$ and $0$, corresponding to the maximal
entanglement state and product state, respectively. However, when we consider
the influence of thermal noise of MO, the entanglement can be demolished.
The black dash line shows the negativity with thermal noise. We
can see that the maximal value can no longer reach 1/2 and it decreases
with increasing of thermal phonon number, which is shown Fig.\ref{entanglement}. This is because that state evolves into mixed state due to the
thermal phonon, and the components of the target state in density matrix
decreases with the increasing of thermal phonon number.

\begin{figure}[htbp]
  \begin{minipage}[t]{\linewidth}
  \includegraphics[width=7cm,angle=0]{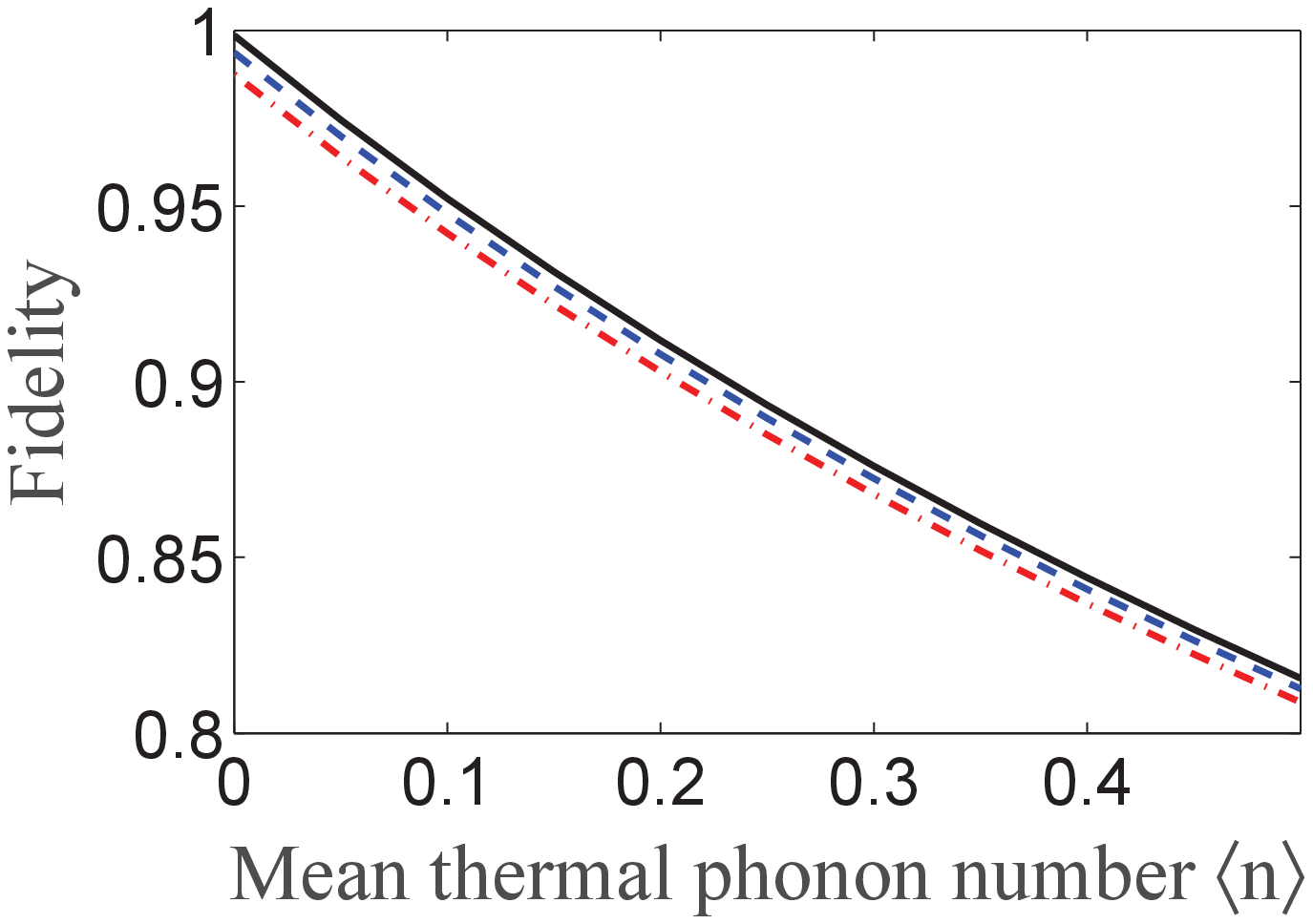}
\end{minipage}
\begin{minipage}[t]{\linewidth}
  \includegraphics[width=7cm,angle=0]{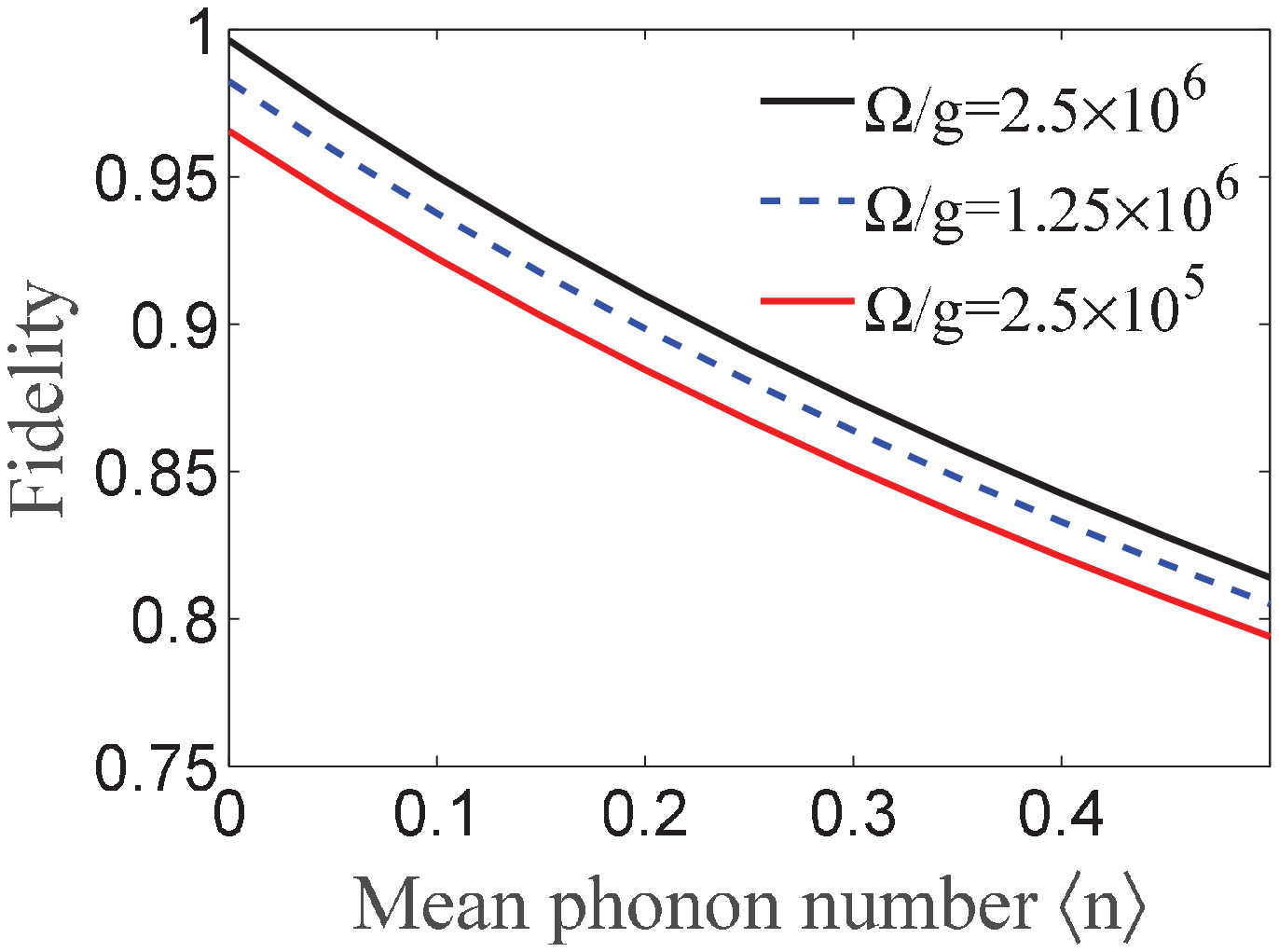}
\end{minipage}
  \caption{(Color online) Fidelity of quantum state transfer and entanglement generation. It is  shown that under different drive strengths, the effects of the MO thermal noise on the quantum state transfer fidelity and entanglement generation fidelity, respectively. The dissipation rate for MO and dephasing rate of NV center are $\gamma/g =5$ and $d/g =50$, respectively.}
  \label{Fidelity}
\end{figure}

The fidelity is of course the essential parameter in quantum information
processing and here we investigate the fidelity of maximal entanglement
state and product state. We define the fidelity between a pure target state $%
|\psi \left( t\right) \rangle $ and mixed state $\rho \left( t\right) $ as $%
F=\sqrt{\langle \psi \left( t\right) |\rho \left( t\right) |\psi \left(
t\right) \rangle }$. That is, the fidelity is equal to the square root of
the overlap between $|\psi \left( t\right) \rangle $ and $\rho \left(
t\right) .$ Due to the dissipation, dephasing rate and thermal noise, the
fidelity decays with time. As Fig.\ref{Fidelity} shows, we describe the fidelity under
different thermal noise and driving strength. Since the thermal phonon can
mix other states into the density matrix, we can see that the fidelity
decays with thermal mean photon number increasing.
However, if we increase the driving strength, the effective coupling between the NV center and
the MO could be enhanced. Therefore, the operation time for both state transfer and the
entanglement generation can be greatly suppressed. As shown in Fig.\ref{Fidelity}, the influence of thermal noise is
also decreased.

\begin{figure}[htbp]
  \centering
  \begin{minipage}[t]{\linewidth}
  \includegraphics[width=7cm,angle=0]{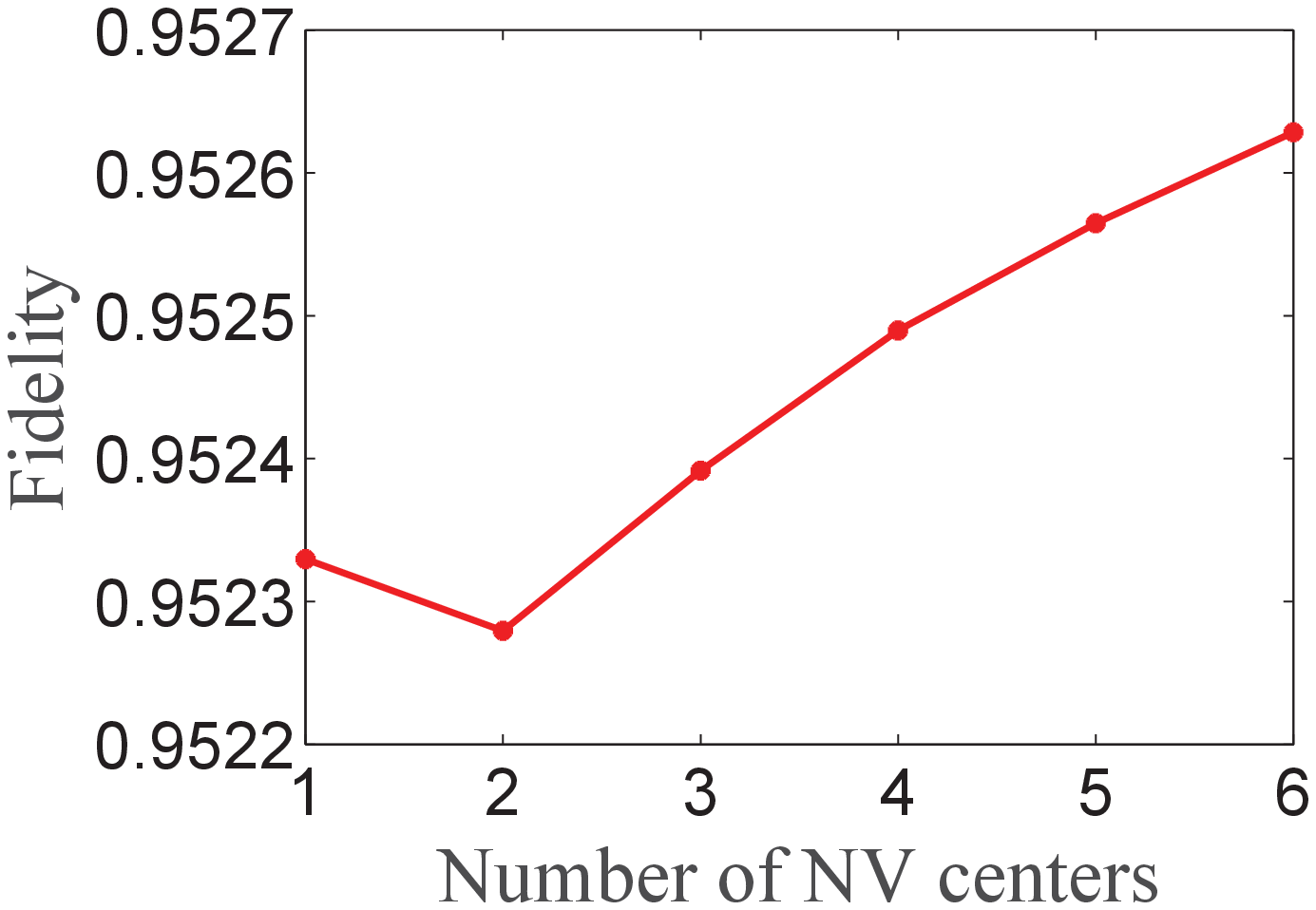}
\end{minipage}
\begin{minipage}[t]{\linewidth}
  \includegraphics[width=7cm,angle=0]{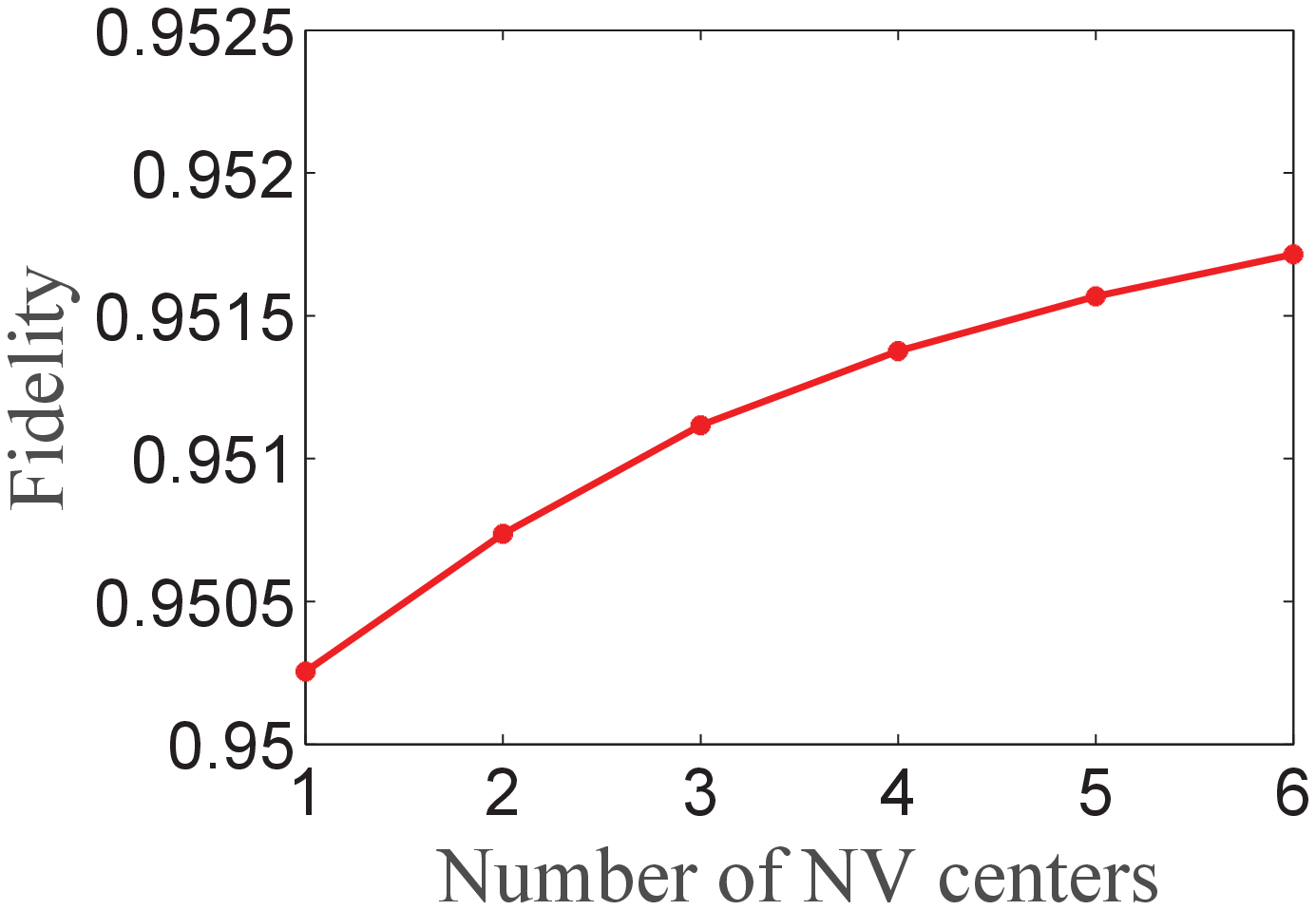}
  \label{(b)}
\end{minipage}
  \caption{(Color online) Fidelity of quantum state transfer between NV centers ensemble and the MO. The relation between the fidelity of (upper figure) quantum state transfer, (lower) or the entanglement generation, and the number of NV centers. The thermal phonon number is $\langle n\rangle =0.1$. The coupling strength is $\Omega/g=2.5\times 10^6$ and the dissipation rate of MO and dephasing rate of NV center are $\gamma/g =5$ and $d/g =50$, respectively.}
\label{FidelityNE}
\end{figure}

As for the interaction of MO and NV centers ensemble, at $t=0$, we suppose that
there are $N-1$ NV centers being in the ground states and then the state
under the evolution of effective Hamiltonian is%
\begin{equation}
\begin{aligned}
|\psi \left( t\right) \rangle=&-\sin \left( g\left\vert \alpha \right\vert
\sqrt{N}t\right) |0,N\rangle \\
 &+\cos \left( g\left\vert \alpha \right\vert
\sqrt{N}t\right) |1,N-1\rangle,
\end{aligned}
\end{equation}%
where $|0,N\rangle $ $\left( |1,N-1\rangle \right) $ denote $N$ $\left(
N-1\right) $ NV centers in the ground state and $0\ \left( 1\right) $ photon
excitation of MO, respectively. From this wave function, we can see that at $%
t=\pi /\left( 4g\left\vert \alpha \right\vert \sqrt{N}\right) $, the maximal
entanglement can be achieved and at $t=\pi /\left( 2g\left\vert \alpha
\right\vert \sqrt{N}\right) $, the state transfer can be realized. At low
temperature experiment, individual relaxation processes can be ignored and
only the intrinsic spin dephasing and phonon dissipation are considered. The
the dynamical process of the system can be given by the master equation \cite{Lukin2013}%
\begin{equation}
\begin{aligned}
\dot{\rho}_{N} =&-i\left[ H_{E},\rho _{N}\right]+\frac{\gamma _{2}}{2}\left( \overline{n}_{T}+1\right) L\left[ b\right]\rho _{N}\\
&+\frac{\gamma _{2}}{2}\overline{n}_{T}L\left[ b^{\dagger }\right]
\rho _{N}+\frac{d }{2}\sum_{i}L\left[ \sigma _{i}^{z}\right] \rho _{N},
\end{aligned}
\end{equation}%
where$n_T =0.1$, $d =50g$ and $\gamma =5g$. Due to the
higher probability of the interaction between NV centers ensemble and MO, the
effective coupling strength has been enhanced and thus, the speed to achieve
the operation goes fast, which lead to the increasing of fidelity directly.
However, more noise could be induced for NV centers ensemble compared with single NV center, which decreases the fidelity.
The competition of two
factors lead to the final results of fidelity for two operations. In
Fig.\ref{FidelityNE} (Upper plot), we can see that at beginning, the effect of noise is more
obvious and then, the effect of enhanced effective coupling strength play
the leader role. Therefore, the fidelity of maximal entanglement decreases
firstly and then rises up latter. But for state transfer as in
Fig.\ref{FidelityNE} (Lower plot), the effect of enhanced effective coupling strength is more
obvious that of noise all the time and thus, the fidelity is monotic
function of the number of NV centers.

\section{Two-mode Squeezing of Mechanical Oscillator}

Besides, the second order interaction can be applied to realize two-mode
squeezing. Keep the design and take an extra mode $c$ into consideration,
with $\omega _{a}-\omega _{c}=\omega _{b}-\omega _{a}$. Mode $a$ resonates
with driver, and through this system, we can realize two-mode squeezing. In
the rotating frame $H_{10}^{\prime }=\omega _{a}a^{\dagger }a+\omega
_{a}b^{\dagger }b+\omega _{a}c^{\dagger }c$, the three modes Hamiltonian has
the form\qquad
\begin{equation}
\begin{aligned}
H_{S0} =&\frac{\omega _{z}}{2}\sigma _{z}+\bigtriangleup b^{\dagger
}b-\bigtriangleup c^{\dagger }c+\frac{\Omega _{1}}{2}\left( a+a^{\dagger
}\right)\\
&+\frac{\Omega _{2}}{2}\left( b+b^{\dagger }\right) +\frac{\Omega
_{2}}{2}\left( c+c^{\dagger }\right),  \\
H_{S1} =&\bigskip \left[ g_{a}a^{\dagger }a+g_{b}b^{\dagger }b+g_{c}c^{\dagger
}c+g_{ab}\left( a^{\dagger }b+b^{\dagger }a\right)\right.\\
&\left. +g_{ac}\left( a^{\dagger
}c+c^{\dagger }a\right) +g_{bc}\left( b^{\dagger }c+c^{\dagger }b\right) %
\right] \sigma _{x},
\end{aligned}
\end{equation}%
where $\bigtriangleup =\omega _{b}-\omega _{a}=\omega _{a}-\omega _{c}\neq
\omega _{z}$. Firstly, we initialize the NV center into the
ground state, and drive the MO to the steady state. At
steady state, the MO moves coherently and make
transformation as $a\rightarrow a+\alpha ,b\rightarrow b+\beta ,c\rightarrow
c+\zeta $, where $\alpha $, $\beta $ and $\zeta $ are amplitude coherent
state. Similar as  Eq. \eqref{amplitude}, we can get $\alpha =-i\Omega _{1}/\gamma _{1},
$ $\beta \simeq -\Omega _{1}/2\triangle $ and $\zeta \simeq \Omega
_{1}/2\triangle $, where $\alpha \gg \beta $, $\zeta $. At this moment, we
turn the NV center into the excited state, and the NV\ center and MO begin
to interact with each other at once. In rotating frame $H_{20}^{\prime }=%
\frac{\omega _{z}}{2}\sigma _{z}+\bigtriangleup b^{\dagger }b-\bigtriangleup
c^{\dagger }c,$ neglecting high frequency and small terms, we have
\begin{equation}
\begin{aligned}
H_{SI}=&\bigskip g\left( \alpha ^{\ast }b\sigma ^{+}e^{i\omega t}+\alpha b^{\dagger
}\sigma ^{-}e^{-i\omega t}\right.\\
&\left. +\alpha ^{\ast }c^{\dagger }\sigma ^{+}e^{i\omega
t}+\alpha c\sigma ^{-}e^{-i\omega t}\right),
\end{aligned}
\end{equation}%
where $\omega =\omega _{z}-\bigtriangleup $ and $g_{ab}=g_{ac}=g.$
Furthermore, we can get the effective squeezing Hamiltonian as%
\begin{equation}
H_{S}=\eta \left( b^{\dagger }b+c^{\dagger }c+bc+c^{\dagger }b^{\dagger
}\right),
\end{equation}%
where $\eta =\left\vert \alpha
\right\vert ^{2}g^{2}\sigma _{z}/\omega .$ As mentioned before, $2\pi
\left\vert \alpha \right\vert /g=10000$, $\omega /g=10^{6}$, thus, we can
get $\eta /g=2500.$ Apply this interaction to the initial state and we will
see
\begin{equation}
|\psi \left( t\right) \rangle _{S}=\exp \left[ -i\eta \left( b^{\dagger
}b+c^{\dagger }c+bc+c^{\dagger }b^{\dagger }\right) t\right] |\psi \left(
0\right) \rangle _{S}.
\end{equation}%
Let $\xi =\eta t$. We define collective creation and annihilation operators
as $d^{\dagger }=\left( b^{\dagger }+e^{i\delta }c^{\dagger }\right) /\sqrt{2%
}$ and $d=\left( b+e^{-i\delta }c\right) /\sqrt{2}$, with $\delta$ the phase of two modes. Based on the collective
operator, the in-phase and in-quadrature components are given by $d_{1}=%
\frac{1}{2}\left( d+d^{\dagger }\right) $ and $d_{2}=\frac{1}{2i}\left(
d-d^{\dagger }\right) $, respectively [21]. The variance of $d_{1}$ in the
two modes squeezed vaccum is%
\begin{equation}
\langle d_{1}^{2}\rangle =\frac{1+2\left( 1-\cos \left( \delta \right)
\right) \xi ^{2}-2\sin \left( \delta \right) \xi }{4}.
\end{equation}%
At space $\left( 0,2\pi \right) $, we can see at $\xi =\frac{\sin \left(
\delta \right) }{2\left[ 1-\cos \left( \delta \right) \right] },$ the result
is
\begin{equation}
\langle d_{1}^{2}\rangle _{\min }=\frac{1-\cos \left( \delta \right) }{8}.
\end{equation}%
From this result, we can see that the minimal value of $\langle
d_{1}^{2}\rangle $ depends on $\delta $. If we choose $\delta =\pi /2$, $%
\langle d_{1}^{2}\rangle =1/8,$ the purpose of squeezing has been realized.

\begin{figure}[htbp]
  \begin{minipage}[t]{\linewidth}
  \includegraphics[width=7cm,angle=0]{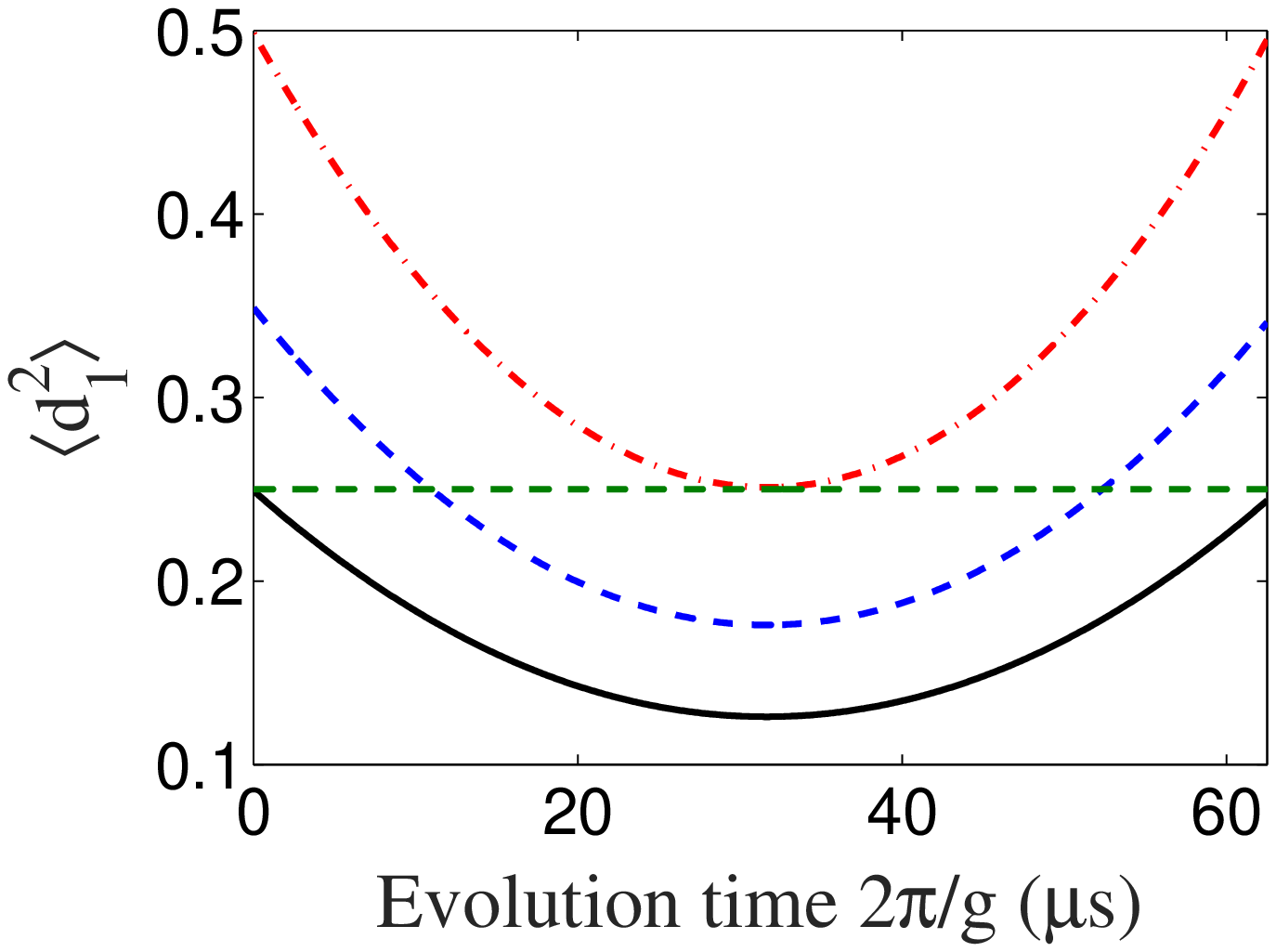}
\end{minipage}
\begin{minipage}[t]{\linewidth}
  \includegraphics[width=7cm,angle=0]{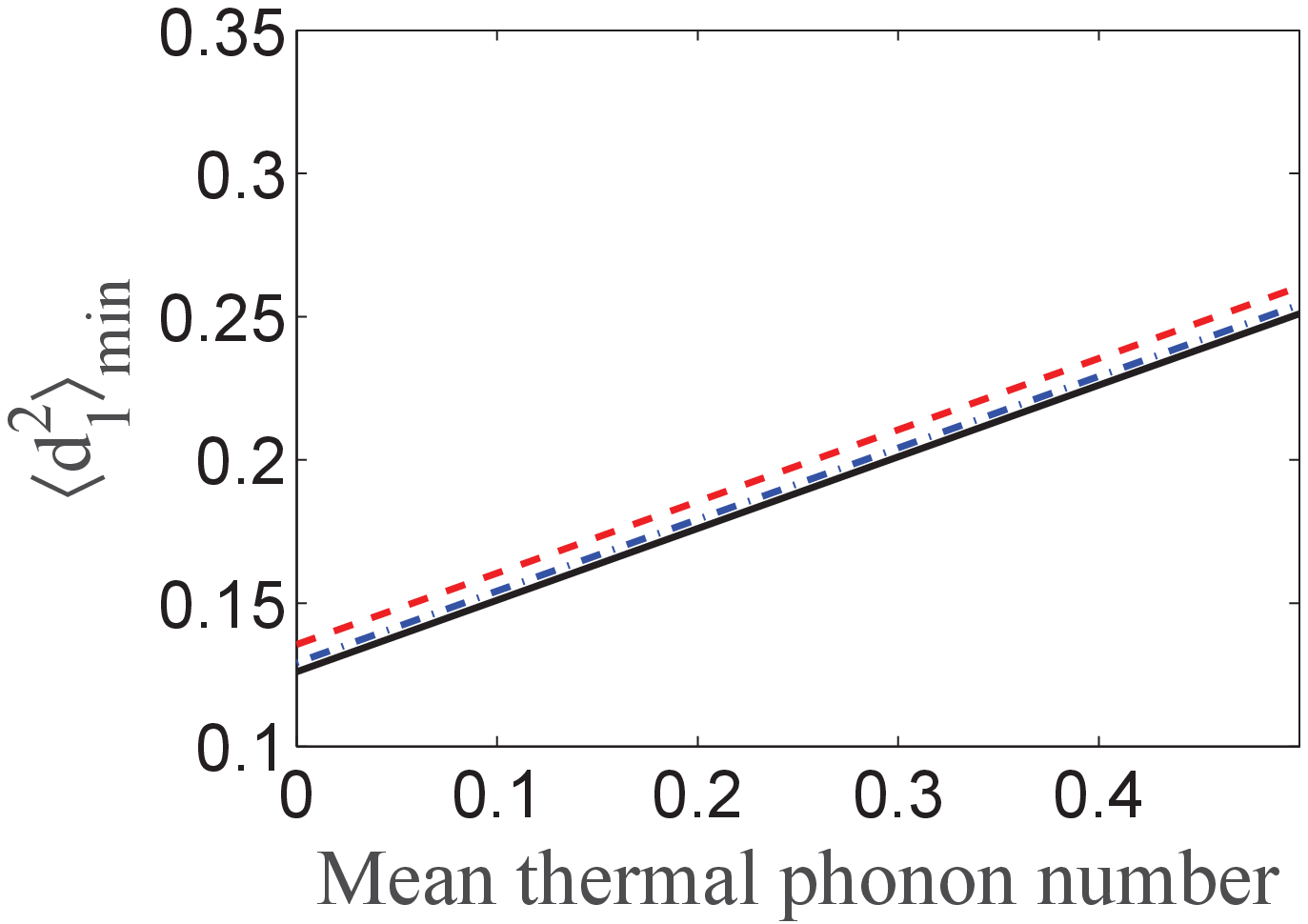}
\end{minipage}
  \caption{(Color online) Squeezing of collective mode $d$. Upper figure shows the dynamic process of $\langle d_1^2\rangle$  with $\langle n\rangle=0,0.2,0.5$. The green dash line shows the critical value $1/4$. The coupling strength is $g/2\pi=5Hz$, the driving strength is $\Omega/g=2.5\times 10^6$ and the dissipation rate of MO is $\gamma_{2}/g =\gamma_{3}/g =\gamma/g =5$. Lower figure shows the relation between the minimal value of $\langle d_1^2\rangle$ and thermal phonon number of MO in three different conditions. The driving strength is $\Omega/g=2.5\times 10^6$, the dissipation rate of MO is $\gamma_{2}/g =\gamma_{3}/g =\gamma/g =5$ for black line, the driving strength is $\Omega/g=2.5\times 10^6$, the dissipation rate of MO is $\gamma_{2}/g =\gamma_{3}/g =\gamma/g =50$ for red dash line and the driving strength is $\Omega/g=1.25\times 10^6$, the dissipation rate of MO is $\gamma_{2}/g =\gamma_{3}/g =\gamma/g =5$ for blue dot dash line.  The dephasing rate of NV center is $d/g =50$ in these two figures.}
  \label{squeezing}
\end{figure}

However, in experiment, the dissipation of the modes should be accounted.
Suppose the mechanical modes have been cooled nearly to the ground state and
the evolution of the state is given by master equation as%
\begin{equation}
\begin{aligned}
\dot{\rho}_{S}\left( t\right)=&-i\left[ H_{S},\rho _{S}\left( t\right) %
\right]+\frac{\gamma _{2}}{2}\left( \overline{n}_{T}+1\right) L\left[ b\right]
\rho _{S}\left( t\right)\\
 &+\frac{\gamma _{2}}{2}\overline{n}_{T}L\left[ b%
\right] \rho _{S}\left( t\right)
+\frac{\gamma _{3}}{2}\left( \overline{n}_{T}+1\right) L\left[ c\right]
\rho _{S}\left( t\right)\\
 &+\frac{\gamma _{3}}{2}\overline{n}_{T}L\left[ c%
\right] \rho _{S}\left( t\right),
\end{aligned}
\end{equation}%
where $L\left[ o\right] \rho =2o\rho o^{\dagger }-o^{\dagger }o\rho -\rho
o^{\dagger }o$ and $\gamma _{2}$, $\gamma _{3}$ are dissipation rate\ of
mode $b$ and $c$, respectively. We can see that because of the collective
correlation, the two-mode squeezing can be realized and the variance of $%
d_{1}$ decreases from $1/4$ to the minimal value $1/8$ when we choose $%
\delta =\pi /2$. However, the correlation can be demolished by thermal noise
and thus, the squeezing is less obvious if more thermal noise is introduced.
As in Fig.\ref{squeezing}, squeezing strength decreases and
finally disappears when thermal phonon number reaches $0.5$. Also, we can
see that the minimal value of the variance of $d_{1}$ increases linearly
with the number of thermal phonon number. The slope rate has nothing to do
with the decay rate of MO and driving strength and thus, we can get that $
\partial \langle d_{1}^{2}\rangle_{\min}/ \partial \rangle n \langle =k $, with $k\simeq 0.25$ for this system.

\section{Discussion and Conclusion}

In conclusion, we have proposed a scheme to realize strong coupling between the MO and NV centers via
the second order magnetic field gradient. We have shown that the effective coupling between can
be greatly enhanced by coherently driving on the MO. We discussed several applications, such as
quantum state transfer, entanglement generation, etc. We also discuss the coupling between NV centers ensemble
and the MO. The thermal noise and the dissipation for both the MO and the NV centers have been discussed.
It is found that high fidelity quantum state transfer and high quality entanglement could be realized for the
present experimental conditions. We also discuss how to generate two-mode squeezing for MO through
coupling to NV center and the external driving. The thermal noise effect on squeezing has been simulated. It is
found that the squeezing could appear  under the thermal phonon number larger than $0$.
We hope that our study could stimulate the further experimental research on the applications of the second order
magnetic field gradient in hybrid quantum systems.

\section*{Acknowledgement}
This work was supported by the National Natural Science Foundation
of China Grant Nos. (61435007, 11175094 , 91221205), the National Basic
Research Program of China (2015CB921002).

\bigskip

\end{document}